\documentclass[11pt]{article}
\usepackage[T1]{fontenc}
\usepackage[utf8]{inputenc}
\usepackage{lmodern}
\usepackage{textcomp}
\usepackage[margin=1in]{geometry}
\usepackage{graphicx}
\usepackage{amsmath,amssymb,amsthm}
\usepackage{cite}
\usepackage{url}
\usepackage{csquotes}
\usepackage{authblk}
\usepackage[hidelinks]{hyperref}
\theoremstyle{plain}
\newtheorem{theorem}{Theorem}
\newtheorem{corollary}{Corollary}
\newtheorem{proposition}{Proposition}
\newtheorem{assumption}{Assumption}
\theoremstyle{definition}
\newtheorem{definition}{Definition}
\newtheorem{criterion}{Criterion}
\theoremstyle{remark}
\newtheorem{remark}{Remark}

\newcommand{\Adv}{\mathrm{Adv}}
\newcommand{\Comb}{\mathsf{Comb}}
\newcommand{\HKDF}{\mathsf{HKDF}}
\newcommand{\ctx}{\mathrm{ctx}}

\title{Correlated Failures in Hybrid Post-Quantum and Quantum Key Distribution: A Formal Security Model}

\author[1]{Mona Alotaibi\thanks{Corresponding author: \href{mailto:ma2315029@qu.edu.qa}{ma2315029@qu.edu.qa}}}
\author[1]{Noora Fetais\thanks{\href{mailto:n.almarri@qu.edu.qa}{n.almarri@qu.edu.qa}}}
\author[2]{Saif Al-Kuwari\thanks{\href{mailto:smalkuwari@hbku.edu.qa}{smalkuwari@hbku.edu.qa}}}
\affil[1]{Department of Computer Science and Engineering, Qatar University, Doha, Qatar}
\affil[2]{Qatar Center for Quantum Computing, College of Science and Engineering, Hamad Bin Khalifa University, Doha, Qatar}
\date{}

\begin{document}
\maketitle

\begin{abstract}
Defense in depth is possible with hybrid key exchange systems that combine post-quantum cryptography and quantum key exchange. Previous research has produced cryptographic security guaranties for a hybrid key exchange system that are at least as robust as those for its best component, but has not examined its defense in depth. This study addresses that gap using a security model. A joint probability distribution represents security concerns for components of the hybrid key exchange system, such as cryptographic assumptions, physical systems, authentication keys, client and server computers, and human users. A key indistinguishability game model reveals the key at each tier if a component is compromised. Proven bounds link the hybrid system's key's maximal distinguishing advantage to the probability of both layers' keys being disclosed up to tiny cryptographic terms. Using a shared-cause dependence parameter limits the security gains from adding a second layer to one bit. With shared causes, total security is limited by -log shared-cause probability and extra security bits from a second layer by $\log_2(1+a_P/c)$, where $a_P$ and $c$ represent PQC-specific and shared-cause compromise probabilities. If the shared-cause probability exceeds the post-quantum layer failure probability, a second layer can only add one bit of protection. Research indicates that a fail-open hybrid of PQC and QKD, with PQC as a backup for QKD, provides no additional security benefits and evaluates the security of the system without using postquantum cryptography. The same approach applies and gives 33 bits of security per block of the QKD protocol for a finite-key parameter of $\varepsilon_Q=10^{-10}$ in the QKD protocol. This is far below the 128-bit security level that might be assumed.

\end{abstract}

\section{Introduction}
\label{sec:introduction}

Two different strategies for creating cryptography that is secure against a quantum computer have been pursued in parallel. Post-quantum cryptography is built using hardness assumptions distinct from those underlying classical cryptography and has been standardized as ML-KEM, ML-DSA, and SLH-DSA \cite{NIST-FIPS203,NIST-FIPS204,NIST-FIPS205}. Quantum key distribution uses physical principles to offer composable security against an adversary with unlimited computing power \cite{bennett2014quantum,ekert1991quantum,quantum_lo_2014,portmann2022security}.

The two technologies have their own security risks. Post-quantum technologies face threats related to new attack vectors and implementation attacks, such as side-channel and fault attacks \cite{ravi2024side,Ravi2024-Blind-Kyber,Krahmer2024-Fault-Dilithium}. QKD technologies face threats related to implementation in physical hardware. Commercial QKD systems have been successfully attacked using detector blinding techniques \cite{lydersen2010hacking}. Hybrid systems incorporating both post-quantum and QKD technologies provide dual protections against attack \cite{dowling2020mickle,bruckner2023muckle,enhanced_garca_2025,hybrid_maduni_2025}. Security challenges for hybrid systems are described in \cite{alotaibi2026survey}; this paper answers one open question from the survey.

The security of constructions such as this can be justified by a combiner theorem that says that if the key derivation function is secure, the hybrid key is pseudorandom if either of the inputs to the key derivation function is pseudorandom \cite{giacon2018kem,bindel2019hybrid,dowling2020mickle,backendal2025kdf}. This security is maintained even if an adversary knows one or both of the inputs to the key derivation function. Similar constructions can be secure even with some correlation between the two inputs, though they do not provide tight security bounds, meaning that while the hybrid construction is guaranteed to be at least as secure as the more-secure input, we cannot say by how much, as layering a quantum input on top of a post-quantum input increases security.

Additional security benefits depend on the failure modes of both layers. In real hybrid systems, there are other connections between the two layers besides key sharing. Existing security measures for QKD systems, such as authentication, must be implemented in hybrid systems with post-quantum technologies. A real-time forgery of the post-quantum signature key used to authenticate both the key exchange and QKD post-processing would also defeat both layers. Other shared resources between layers of a hybrid system include microarchitectural vulnerabilities of their underlying hardware platforms, such as GoFetch \cite{GoFetch2024}, and common manufacturing and operational supply chains. Engineering research shows that shared resources reduce the value of redundancy \cite{avizienis2004basic,knight1986experimental,eckhardt1985theoretical,littlewood2004redundancy}. This paper provides a formal analysis of security benefits of hybrid PQC-QKD key establishment systems.

\subsection{Contributions}

\begin{enumerate}
\item \emph{A security experiment with externally determined disclosure} (Section~\ref{sec:model}). Individual components are disclosed according to some joint distribution, and each layer's secret key is disclosed if and only if a related component is disclosed. Applications include cryptographic assumption failures, user device compromises, long-running system compromises, and operating environment sharing among users and hosts.

\item \emph{Exact security bound} (Theorem~\ref{thm:bridge}). The gain to attackers is at least $(1-p_{\bot})\pi_{11}(1-2^{-\ell})$ and at most $\pi_{11}+\max(\eta_P,\eta_Q)$, where \(\pi_{11}\) is the probability that both underlying keys are broken, and \(\eta_P\) and \(\eta_Q\) are the security levels of the underlying keys and combiner, respectively. Breaking both underlying keys significantly weakens the hybrid key.

\item \emph{Two-sided margin bounds} (Proposition~\ref{prop:margin}). For measures of dependence with parameter $\rho$, rescaled to accommodate negative dependence, the two-sided margin over the stronger component is between $\min(n,-\log_2\rho)-1$ and $\min(n,-\log_2\rho)$.

\item \emph{Common-cause cap} (Theorem~\ref{thm:cc}). The hybrid offers no more than $-\log_2 c$ bits of security, where $c$ is the common-cause probability, and its margin over PQC alone is at most $\log_2(1+a_P/c)$ bits. Multiple independent common causes further reduce security by no more than another $\log_2 k$ bits beyond the best common cause (Corollary~\ref{cor:multi}).

\item \emph{Combination modes} (Section~\ref{sec:arch}): Serial, parallel, and salted HKDF key combinations are secure under some assumptions about the security of HMAC (Proposition~\ref{prop:modes}). A fallback scheme using postquantum algorithms in case QKD fails offers no benefits over the postquantum algorithms alone (Proposition~\ref{prop:fallback}).

\item \emph{Exact security criterion} (Proposition~\ref{prop:degrade}). In the worst-case quantum setting, the entire system provides $2^{-\lambda_T}$ security if and only if the QKD system alone provides $2^{-\lambda_T}$ security and the accuracy with which the QKD system's security can be verified is limited by the finite-key security parameter.
\end{enumerate}

This paper is theoretical in nature. Proofs are provided in Appendix~\ref{app:proofs}. Numerical examples in Section~\ref{sec:numerical} are computed using closed-form solutions. Online Resource~1 reproduces the numerical examples.

\section{Related Work}
\label{sec:related}

\subsection{Combiners and Hybrid Key Establishment}

Robust combiners offer security as long as at least one component of the combiner is secure \cite{harnik2005robust}. Giacon, Heuer, and Poettering \cite{giacon2018kem} discuss conditions and results for combined key encapsulation methods with IND-CCA security, including an XOR combination of key encapsulation methods that requires binding the ciphertexts into key derivation. Results related to hybrid key encapsulation and authenticated key exchange are given by Bindel et al.\ \cite{bindel2019hybrid}. Krawczyk's HKDF \cite{krawczyk2010hkdf} is a common key derivation technique; Backendal et al.\ \cite{backendal2023hmac} investigate assumptions about dual pseudorandom functions, including HMAC, commonly used in hybrid constructions. Results and techniques for multi-input key derivation, including inputs that may be correlated, are discussed by Backendal et al.\ \cite{backendal2025kdf} and applied to hybrid key exchange.

Hybrid QKD protocols have been studied in provable security models by Dowling, Hansen, and Paterson \cite{dowling2020mickle}. In the HAKE hybrid protocol, the adversary can learn any combination of classical and post-quantum and QKD secret information. Other hybrid protocols with digital signature, and KEM-based authentication include Muckle+ \cite{bruckner2023muckle} and Muckle\# \cite{battarbee2024muckle}. Mosca, Stebila, and Ustao\u{g}lu \cite{mosca2013qkd} show that classical authentication technology can be used to protect a QKD protocol without affecting its long-term security as long as the QKD protocol continues to use classical authentication technology. H\"ovelmanns et al.\ \cite{hovelmanns2025oracles} discuss key compromise impersonation and other attacks on QKD hybrid protocols involving use of QKD keys and suggest ways to model the QKD component of these protocols. Clermont and Henrich \cite{clermont2026best} suggest eliminating signatures from QKD-PQC hybrid authentication protocols. Other relevant work on analyzing and proving hybrid protocols includes \cite{chen2025hybrid,raj2025hybrid,enhanced_garca_2025,hybrid_maduni_2025}.

\subsection{What the Present Model Adds}

The frameworks described above are worst-case frameworks that prove the security of the hybrid key under any compromise pattern in which at least one of the layers is not compromised, including arbitrary correlation across layers. The frameworks do not help determine the likelihood of different compromise patterns or quantify the benefits to security of adding a second layer. We extend the worst-case frameworks described above (Theorem~\ref{thm:bridge}) to consider a distribution over compromise patterns. Theorem~\ref{thm:bridge} is narrowly scoped and intended primarily as a tool to connect the probabilities of distinguishing attacks and compromise patterns to allow formal rather than informal statements of security results from Sections~\ref{sec:dependence} and~\ref{sec:cc} in terms of risk. Our framework differs from the correlated-input KDF framework of \cite{backendal2025kdf} in that we consider correlations in compromise events. Deployment considerations related to this work are discussed in \cite{aquina2025critical}.

\subsection{Dependent Failure in Redundant Systems}
\label{sec:ccf}

The taxonomy of dependable and secure systems \cite{avizienis2004basic} defines concepts of independent and common-cause failures. Knight and Leveson \cite{knight1986experimental} provide evidence of high rates of common-cause failure in software systems, and Eckhardt and Lee \cite{eckhardt1985theoretical} offer explanations. Littlewood and Strigini \cite{littlewood2004redundancy} have qualitatively extended these results in the security domain. Other work on probabilistic risk assessment includes parametric models of common cause, such as the beta factor model \cite{mosleh1998guidelines}, and statistical models of common cause, such as those of Marshall and Olkin \cite{marshall1967multivariate}. The dependence parameter described in Section~\ref{sec:dependence} can be interpreted as a normalized measure of dependence between events, equivalent to a Fr\'echet mixture of independent and comonotonic dependence \cite{nelsen2006copulas}; it is closely related to the beta factor model when the two events have the same distribution. Our primary contribution is not a new dependence model but a connection between that model and a cryptographic experiment, along with some bounds.

\section{Model}
\label{sec:model}

\subsection{System}

Two parties run two key-agreement protocols in one session. The PQ key-agreement protocol generates a key $K_P\in\{0,1\}^{\ell}$ using a key encapsulation mechanism. The QKD key-agreement protocol, which implements a discrete-variable QKD protocol and relies on an authenticated classical channel for classical postprocessing steps including sifting, error correction, and privacy amplification, generates a key $K_Q\in\{0,1\}^{\ell}$ or aborts. The two key-agreement protocols are run independently of one another and with independent randomness. Let $\ctx$ be the transcript of the session, including the KEM ciphertext and a QKD key identifier \cite{hovelmanns2025oracles}. A combiner uses inputs $(K_P,K_Q,\ctx)$ to generate a session key $K\in\{0,1\}^{\ell}$; we use $\ell=256$ throughout. The session produces no key if either key-agreement protocol aborts. See Section~\ref{sec:arch} for information about fallback alternatives.

\subsection{Components and Exposure}

Components are compromised in Table~\ref{tab:components}. Components are anything that, if compromised during the key's protected lifetime, could expose a layer's key. Components include cryptographic assumptions, devices, authentication keys, hosts, and users. For each component \(k\) in a finite set \(\mathcal{K}\), there is a subset \(X(k)\subseteq\{P,Q\}\) of layers that would be exposed by compromising \(k\). For a subset \(S\subseteq\mathcal{K}\) of compromised components, define events
\begin{align*}
E_P(S) &= [\exists k\in S:\ P\in X(k)],\\
E_Q(S) &= [\exists k\in S:\ Q\in X(k)].
\end{align*}

\begin{table}[t]
\caption{Components and the layers whose keys they expose}
\label{tab:components}
\centering
\small
\setlength{\tabcolsep}{3pt}
\begin{tabular}{@{}p{2.6cm}cp{3.6cm}@{}}
\hline\noalign{\smallskip}
\textbf{Component} & \textbf{Exposes} & \textbf{Examples} \\
\hline\noalign{\smallskip}
KEM (\texttt{kem}) & $P$ & Cryptanalysis of the lattice assumption (including after the session); side channels \\
QKD device (\texttt{qkd}) & $Q$ & Detector blinding, source flaws; trusted relays \\
Shared authentication (\texttt{auth}) & $P,Q$ & Real-time compromise of the signing key that authenticates both the KEM exchange and QKD post-processing\textsuperscript{a} \\
Shared host (\texttt{host}) & $P,Q$ & Microarchitectural attack, implant \\
Shared operator (\texttt{op}) & $P,Q$ & Key-management, credential or supply-chain compromise \\
\hline
\end{tabular}

\smallskip
\raggedright\footnotesize\textsuperscript{a}A break of the signature scheme after the session exposes nothing \cite{mosca2013qkd}. With separate authentication keys, or pre-shared-key authentication of QKD post-processing, \texttt{auth} splits into components exposing $P$ and $Q$ separately
\end{table}

\begin{assumption}[Exogenous compromise]
\label{asm:exo}
The set $S$ of compromised components is chosen independently according to a distribution $\mathcal{D}$ over subsets of $\mathcal{K}$ and is independent of the challenge bit and the random bits used by both layers.
\end{assumption}

Assumption~\ref{asm:exo} leaves open an arbitrary relationship between components; their relationship is governed by $\mathcal{D}$. Assumption~\ref{asm:exo} excludes adversaries who can make compromise depend on the key values themselves. We use
\begin{gather*}
\pi_{xy} = \Pr_{S\leftarrow\mathcal{D}}\big[\mathbf{1}_{E_P(S)}=x,\ \mathbf{1}_{E_Q(S)}=y\big],\\
p=\pi_{10}+\pi_{11},\qquad q=\pi_{01}+\pi_{11}.
\end{gather*}

\subsection{Security Experiment}

\begin{definition}[Hybrid key indistinguishability]
\label{def:exp}
In the experiment $\mathrm{Exp}^{\mathrm{hyb}}_{\Pi,\mathcal{D}}(\mathcal{A})$, the challenger samples $S\leftarrow\mathcal{D}$ and $b\leftarrow\{0,1\}$ and gives $S$ to $\mathcal{A}$. Protocol $\Pi$ is executed. The attacker $\mathcal{A}$ has full control over the classical channel and can manipulate the classical channel only if $\texttt{auth}\in S$. If the session is aborted, $\mathcal{A}$ receives $\bot$. Otherwise, $\mathcal{A}$ receives $\ctx$, quantum side information, $K_P$ if $E_P(S)$, $K_Q$ if $E_Q(S)$, and $K_b$, where $K_0=K$ and $K_1$ is uniform in $\{0,1\}^\ell$. $\mathcal{A}$ outputs a bit $b'$.
\[
\Adv^{\mathrm{hyb}}_{\Pi,\mathcal{D}}(\mathcal{A}) = \big|\Pr[b'=1\mid b=0]-\Pr[b'=1\mid b=1]\big|.
\]
\end{definition}

Disclosing the full key for an exposed layer is the best option to cover any partial leakage. Repeating the experiment using $K := \mathsf{KDF}(K_P,\ctx)$ (or $K_Q$) defines a standalone PQC (or QKD) deployment exposed to the same components.

\subsection{Component Assumptions}

\begin{assumption}[Layers and combiner]
\label{asm:comp}
\begin{enumerate}
\item[(a)] \emph{PQC.} For all polynomial-time distinguishers active at time $t$ and all polynomial-time-computable auxiliary inputs that are independent of $K_P$ given the PQC transcript, the key $K_P$ is $\delta_P(t)$-indistinguishable from uniformly random given the PQC transcript and auxiliary input.
\item[(b)] \emph{QKD.} If the classical postprocessing channel is authentic and the protocol does not abort, the joint state of $K_Q$ and all other systems is $\varepsilon_Q$-close in trace distance to a state with $K_Q$ replaced by a uniformly random and independent string \cite{portmann2022security,tomamichel2012tight}.
\item[(c)] \emph{Combiner.} For a uniformly distributed and secret $K_P$, the value of $\Comb(K_P,k_Q,\ctx)$ is $\delta_{c,P}(t)$-indistinguishable from uniform for all $k_Q$ and $\ctx$ known to the distinguisher. Similarly, for a uniformly distributed and secret $K_Q$, the value of $\Comb(k_P,K_Q,\ctx)$ is $\delta_{c,Q}(t)$-indistinguishable from uniform for all $k_P$ known to the distinguisher.
\end{enumerate}
\end{assumption}

We define $\eta_P=\delta_P(t')+\delta_{c,P}(t')$ and $\eta_Q=\varepsilon_Q+\delta_{c,Q}(t')$, where $t'$ is greater than $t$ by the cost of simulating honest behavior on one layer. $\eta_P$ and $\eta_Q$ are called the \emph{cryptographic residuals}. These quantities are usually negligible in ordinary analyses. The only non-negligible quantity in ordinary analyses is the exposure probability.

\subsection{Risk-Level Security}

Probabilities are stated in terms of the key's lifetime of protection. Probabilities are measured in bits as $\lambda = -\log_2(\cdot)$. Notation:
\[
\lambda_H=-\log_2\pi_{11},\quad\lambda_P=-\log_2 p,\quad\lambda_Q=-\log_2 q,
\]
with $m=\max(\lambda_P,\lambda_Q)$ and $n=\min(\lambda_P,\lambda_Q)$. These are security levels. They are distinct from the bit security of an implementation of a primitive \cite{micciancio2018bit}. Bit security of an implementation impacts only the residual error terms $\delta_P(t)$. For instance, $\lambda_P$ is not the NIST category-3 security level of ML-KEM-768 but the probability of failure of the ML-KEM-768 lattice problem or any shared sub-component in the implementation over the lifetime of the ML-KEM-768 primitive. No cryptanalysis provides an estimate of failure probability.

\section{Hybrid Security under Exogenous Compromise}
\label{sec:security}

\begin{theorem}[Joint exposure determines hybrid security]
\label{thm:bridge}
Under Assumptions~\ref{asm:exo} and~\ref{asm:comp}, every adversary $\mathcal{A}$ running in time $t$ satisfies
\begin{align}
\Adv^{\mathrm{hyb}}_{\Pi,\mathcal{D}}(\mathcal{A})
&\le \pi_{11} + \pi_{10}\,\eta_Q + \pi_{01}\,\eta_P + \pi_{00}\min(\eta_P,\eta_Q) \notag\\
&\le \pi_{11} + \max(\eta_P,\eta_Q).
\label{eq:bridge}
\end{align}
Conversely, there is a passive adversary $\mathcal{A}^*$, running in the time of one evaluation of $\Comb$, with
\[
\Adv^{\mathrm{hyb}}_{\Pi,\mathcal{D}}(\mathcal{A}^*) \ge (1-p_\bot)\,\pi_{11}\,(1-2^{-\ell}),
\]
where $p_\bot$ is the abort probability of an honest execution.
\end{theorem}

The theorem provides two security measures used elsewhere: the probability that an adversary learns both keys in the hybrid system and the cryptographic residuals that remain if the adversary does not learn both keys. When the latter measure is small compared to $\pi_{11}$, an adversary's advantage $\lambda_H$ in the hybrid system is practically the same as the security of the hybrid key, differing by at most one bit:

\begin{corollary}
\label{cor:exact}
If $\max(\eta_P,\eta_Q)\le\pi_{11}$, the optimal advantage $\Adv^*$ satisfies
$\lambda_H - 1 \le -\log_2\Adv^* \le \lambda_H - \log_2\big((1-p_\bot)(1-2^{-\ell})\big)$.
The same arguments that apply to single deployments show that $\lambda_P$ and $\lambda_Q$ are within the same range of their security levels.
\end{corollary}

\subsection{Gap Between External Threat and Defense-in-Depth}
\label{sec:dependence}

The metric for choosing whether to deploy is the difference between total output when deploying both and output from the stronger standalone deployment.

\begin{definition}[Margin]
\label{def:margin}
The margin is\\ $\Delta = \lambda_H - m = \log_2\big(\min(p,q)/\pi_{11}\big)$.
\end{definition}

Given marginal distributions  $p$ and $q$, the joint value $\pi_{11}$ is bounded by Fr\'echet interval $[\max(0,p+q-1),\ \min(p,q)]$.$\pi_{11}$ is reported in normalized units based on the Fr\'echet bounds.

\begin{definition}[Compromise dependence]
\label{def:rho}
For $0<p,q<1$, the compromise dependence is
\[
\rho = \frac{\pi_{11}-pq}{\min(p,q)-pq}.
\]
equivalently $\pi_{11}=(1-\rho)pq+\rho\min(p,q)$.
\end{definition}

\begin{proposition}[Range of $\rho$]
\label{prop:range}
Let $p+q\le 1$. Then the correlation parameter $\rho$ ranges over the interval $[\rho_{\min},1]$, where $\rho_{\min}=-\max(p,q)/(1-\max(p,q))$. Independence occurs when correlation is zero. Maximum correlation, comonotonicity, or co-increasing dependence occurs when exposure in the less likely layer fully predicts exposure in the other layer. Minimum correlation occurs when the two layers are mutually exclusive, with $\pi_{11}=0$. Every level of correlation between minimum and maximum can be attained by some joint distribution.
\end{proposition}

Negative risk values are possible: if the attacker has a fixed budget that allows attacking either layer for the same cost, the events of compromising both layers are nearly mutually exclusive. If nonnegative, $\rho\approx\Pr[\text{more likely layer exposed}\mid\text{less likely layer exposed}]$ for nonnegative $\rho$; risk measurement is subject to directional bias, but not in the way $\rho$ is calculated, it depends on layer marginals (Remark~\ref{rem:rhocc}).

\begin{proposition}[Two-sided margin bound]
\label{prop:margin}
With $n=\min(\lambda_P,\lambda_Q)$,
\begin{equation}
\Delta = -\log_2\!\big[2^{-n}+\rho\,(1-2^{-n})\big].
\label{eq:margin}
\end{equation}
For $\rho\in[0,1]$, $\ \min(n,-\log_2\rho)-1\ \le\ \Delta\ \le\ \min(n,-\log_2\rho)$. For $\rho<0$, $\Delta>n$.
\end{proposition}

Thus, the secrecy advantage of a hybrid key establishment scheme over its better component scheme is no more than $-\log_2\rho$ bits and no less than $-\log_2(\rho) - 1$ bits for $-\log_2\rho\le n$. Further improvements to the better component will not improve the secrecy advantage of the hybrid scheme beyond some limit for $\rho\ge 2^{-n}$. The multiversion software saturation phenomenon \cite{knight1986experimental,eckhardt1985theoretical} applies to hybrid key establishment.

\begin{proposition}[Quantum floor]
\label{prop:floor}
For all $\mathcal{D}$, $\lambda_H\ge m\ge\lambda_Q$. When the PQC layer is unfolded consistently ($p=1$), $\lambda_H=\lambda_Q$.
\end{proposition}

The floor is immediate, $\pi_{11}\le q$, and holds under any dependence. Its content lies in what $q$ contains. Under harvest-now-decrypt-later, a future break of the KEM is an exposure of $P$ only, so the floor applies. A break of the authentication scheme during execution is an exposure of $Q$ as well (Table~\ref{tab:components}), and is part of $q$.

\section{Common-Cause Compromise}
\label{sec:cc}

The parameter $\rho$ indicates dependence, not the cause of dependence. The models listed in Table~\ref{tab:components} are examples of common-component models. In these models, common failures cause simultaneous failures in both systems. Common-component models are also referred to as common-shock models \cite{marshall1967multivariate}.

\begin{theorem}[Common-cause limit]
\label{thm:cc}
Let $E_P=A_P\cup C$ and $E_Q=A_Q\cup C$ with $A_P$, $A_Q$, $C$ independent events of probabilities $a_P$, $a_Q$, $c$, where $C$ is an attack on the shared component. Then $\pi_{11}=c+(1-c)a_Pa_Q$.
\begin{enumerate}
\item[(i)] $\min\!\big(-\log_2 c,\ -\log_2(a_Pa_Q)\big)-1\ \le\ \lambda_H\ \le\ -\log_2 c$;
\item[(ii)] for $c>0$, combination PQC has greater value than standalone PQC.
\[
\Delta_P := \lambda_H-\lambda_P \le \log_2\!\left(1+\frac{a_P}{c}\right),
\]
and symmetrically, $\Delta_Q:=\lambda_H-\lambda_Q\le\log_2(1+a_Q/c)$.
\end{enumerate}
In particular, when $c\ge a_P$, the combined system exceeds standalone PQC by no more than one bit.
\end{theorem}

Part (ii) is the applied contribution of the paper. Using QKD along with a post-quantum security system provides only minimal additional security benefits compared to using the post-quantum system alone, based on the risk of compromise of the post-quantum system and the risk of simultaneous failure of both the QKD and post-quantum systems. Shared authentication keys and shared hosts or operators with security no greater than the lattice-based assumptions of the post-quantum system cap the security benefit of the QKD system at one bit, independent of the finite-key parameters of the QKD system.

\begin{corollary}[Several common causes]
\label{cor:multi}
If $C=\bigcup_{i=1}^{k}C_i$ and the $C_i$ are independent events with probabilities $c_i$, then $c=1-\prod_i(1-c_i)$ and $\max_i c_i\le c\le\sum_i c_i\le k\max_i c_i$. Therefore, $\lambda_H\le-\log_2\max_i c_i$, and the effect on $\lambda_H$ of removing all but one common cause of maximum probability is less than or equal to $\log_2 k$ bits.
\end{corollary}

Separation strategies should target the largest shared factor. Other shared factors can be separated using at most log base $\log_2 k$ bits.

\begin{remark}[$\rho$ is not a feature of the mechanism]
\label{rem:rhocc}
Under the model of Theorem~\ref{thm:cc}, with small probabilities, $\rho\approx c/(\min(a_P,a_Q)+c)$. One shared feature may lead to different values of $\rho$ for different probabilities at different layers. Keeping $\rho$ fixed while changing probabilities at one layer is not the same as keeping the shared feature fixed. The next theorem keeps $c$ fixed. 
\end{remark}

\begin{proposition}[Persistence in the face of degradation of the PQC layer]
\label{prop:degrade}
Under the model of Theorem~\ref{thm:cc}, with fixed parameters $a_Q > 0$ and $c$, and increasing $a_P$ (due to improvements in cryptanalysis of the KEM), the target is $\pi_{11}\le 2^{-\lambda_T}$. $\pi_{11}$ is a nondecreasing function of $a_P$, and the target is met if and only if

\[
c\le 2^{-\lambda_T}\quad\text{and}\quad a_P\le a_P^*:=\frac{2^{-\lambda_T}-c}{(1-c)\,a_Q}.
\]

The target is still secure if the PQC layer is fully broken ($a_P=1$) if and only if $q\le 2^{-\lambda_T}$, or $\lambda_Q\ge\lambda_T$.
\end{proposition}

The condition $c\le2^{-\lambda_T}$ applies to any layered approach, regardless of the security levels of individual layers. Advantages in security from one layer cannot compensate for a higher probability of a shared cause in the layered construction than is permitted. The condition applies to stacked constructions as well as any other approach involving the same shared component.

\section{Key Combination Modes}
\label{sec:arch}

Write $\HKDF(\mathrm{IKM};\mathrm{salt},\mathrm{info})$ to denote HKDF \cite{krawczyk2010hkdf}. Here, $\mathrm{salt}=0$ denotes the default all-zero salt. Three combination modes are in use:
\begin{align}
\Comb_{\mathsf{ser}} &= \HKDF(K_P\,\|\,K_Q;\ 0,\ \ctx), \label{eq:ser}\\
\Comb_{\mathsf{par}} &= \HKDF(K_P;\ 0,\ \ctx\|\mathtt{p}) \notag\\
 &\quad\oplus \HKDF(K_Q;\ 0,\ \ctx\|\mathtt{q}), \label{eq:par}\\
\Comb_{\mathsf{salt}} &= \HKDF(K_Q;\ K_P,\ \ctx). \label{eq:salt}
\end{align}
Concatenation \eqref{eq:ser} is the pattern of hybrid key exchange in TLS and of most of the literature \cite{bindel2019hybrid,backendal2025kdf}; \eqref{eq:par} is the XOR combiner \cite{giacon2018kem}; \eqref{eq:salt} chains one secret into the salt of the other's extraction, as in the TLS~1.3 key schedule.

\begin{proposition}[Soundness of three modes]
\label{prop:modes}
Each of \eqref{eq:ser}--\eqref{eq:salt} satisfies Assumption~\ref{asm:comp}(c), with $\delta_{c,P}$ and $\delta_{c,Q}$ bounded by the advantage against the following properties of HMAC, plus the PRF advantage of HKDF-Expand:
\begin{itemize}
\item \eqref{eq:ser}: extraction from a key input with one secret and uniform component and one known component (dual-PRF or multi-input KDF property \cite{backendal2023hmac,backendal2025kdf});
\item \eqref{eq:par}: extraction from a uniform secret input with a publicly known salt \cite{krawczyk2010hkdf}; applied separately to each branch;
\item \eqref{eq:salt}: Pseudorandomness of HMAC with key  $K_P$ chosen uniformly at random on the $P$ side, extraction with known salt \cite{krawczyk2010hkdf} on the $Q$ side.
\end{itemize}
Theorem~\ref{thm:bridge} therefore applies to all three graphs.
\end{proposition}

In particular, adding $K_P$ to the salt does not affect the security analysis: the salt for HKDF-Extract is used as the HMAC key input, and a uniformly random HMAC key produces a pseudorandom HMAC output, even if $K_Q$ is known. Differences between the three schemes relate to different assumptions about HMAC, not different analyses of Theorem~\ref{thm:bridge}. Two caveats: To support the parallel scheme, we assume independence of keys at the two layers. Adversaries who can control one layer's input as a function of the other are ruled out by Assumption~\ref{asm:exo} and considered in \cite{backendal2025kdf}. All three schemes require that $\ctx$ bind the ciphertext for the KEM \cite{giacon2018kem} and the QKD key identifier \cite{hovelmanns2025oracles}.

\begin{remark}[Direct use is not mixing]
\label{rem:direct}
An integration in which the session key is $K_Q$ itself and PQC for authentication in the post-processing step of QKD is not a combiner because the PQC key is not used in key generation. Security for this integration is equivalent to that of QKD using only PQC for authentication, $\lambda=\lambda_Q$, and the shared authentication key is a component exposing $Q$ by itself.
\end{remark}

A hybrid approach using a PQC key as a backup if QKD doesn't work can provide availability. QKD can abort under the adversary's control.

\begin{proposition}[Fallback option is costly]
\label{prop:fallback}
Let $\Pi_{\mathsf{fb}}$ be the postclassical protocol in which, on successful QKD, key $K=\Comb(K_P,K_Q,\ctx)$ is established, and on failed QKD, key $K=\HKDF(K_P;0,\ctx\|\mathtt{fb})$ is established. Let $\Pi_P$ be the pure PQC protocol in which $K=\HKDF(K_P;0,\ctx\|\mathtt{fb})$ is established. Adversaries $\mathcal{A}$ and $\mathcal{B}$ attacking $\Pi_{\mathsf{fb}}$ and $\Pi_P$, respectively, have the same $\mathcal{D}$ oracle. For every adversary $\mathcal{B}$ attacking $\Pi_P$, there is an adversary $\mathcal{A}$ attacking $\Pi_{\mathsf{fb}}$ with the same running time as $\mathcal{B}$, plus the time needed to block the quantum channel, such that
\[
\Adv^{\mathrm{hyb}}_{\Pi_{\mathsf{fb}},\mathcal{D}}(\mathcal{A}) = \Adv_{\Pi_P,\mathcal{D}}(\mathcal{B}).
\]
Specifically, the best attack on $\Pi_{\mathsf{fb}}$ has advantage at least $p\,(1-2^{-\ell})$ times the PQC layer's own completion probability, and combining the schemes provides no other security advantages over using only PQC.
\end{proposition}

Blocking or eavesdropping on quantum communication channels is within the capabilities of any attacker considered when analyzing QKD security and will lead to failure of a QKD protocol without compromising security. When providing security protections at multiple layers, a failed QKD protocol run should be treated as a failed session at the cryptographic layer and reported to the application layer, rather than silently using the post-quantum key.

\section{Analytical Illustration}
\label{sec:numerical}

All numbers in this section are obtained from closed-form formulas and verified in Online Resource 1, which numerically verifies the bounds of Propositions~\ref{prop:margin} and~\ref{prop:degrade} and Theorem~\ref{thm:cc} over a range of parameter values. Probability estimates in this section are given for illustration purposes and should not be interpreted as empirical results.

\subsection{Adjusting for Quantum Effects}

The finite-size security parameter $\varepsilon_Q$ provides a bound on the leftover entropy $\eta_Q$ quantified in Theorem~\ref{thm:bridge}. The finite-size security parameter corresponds to each block of key and is often chosen to be about $10^{-10}$ in theoretical analyses \cite{tomamichel2012tight} and 33 bits per block in practical applications of QKD. Lower values of $\varepsilon_Q$ result in lower secret key rates. A secret key rate established by Proposition~\ref{prop:floor} cannot exceed $-\log_2(q+\varepsilon_Q)$, or 33 bits per block, ignoring the possibility of device compromise. Asserting more than 100 information-theoretic bits per protocol run requires $\varepsilon_Q$ less than $10^{-10}$. We report results for $n=33$ herein and, for comparison, $n=100$ to align with the literature.

\subsection{Margin and Dependence}

\begin{figure}[t]
\centering
\includegraphics[width=\columnwidth]{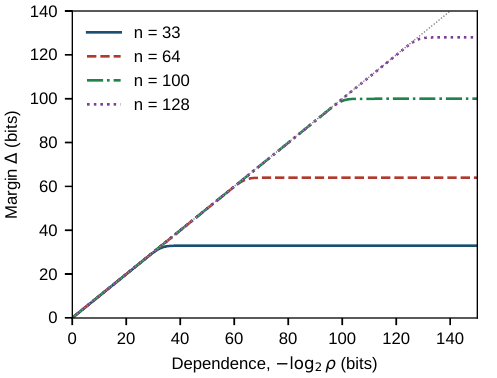}
\caption{Margin $\Delta$ against compromise dependence $-\log_2\rho$ for $n=33$ (solid), $64$ (dashed), $100$ (dash-dot) and $128$ (dotted). The thin dotted diagonal is the envelope $\Delta=-\log_2\rho$. Each curve follows the envelope until it reaches $n$}
\label{fig:margin}
\end{figure}

Table~\ref{tab:margin} gives values for \eqref{eq:margin}. The margin for $n$ is $n$ when $\rho\ll2^{-n}$ and $-\log_2\rho$ for all $n$. For $n=33$, the margin changes to $-\log_2\rho$ when $\rho\approx2^{-33}$. Figure~\ref{fig:margin} gives similar results for four values of $n$.

\begin{table}[t]
\caption{Margin $\Delta$ (bits) over the stronger component as a function of the compromise dependence $\rho$, from \eqref{eq:margin}}
\label{tab:margin}
\centering
\small
\begin{tabular}{@{}lrr@{}}
\hline\noalign{\smallskip}
$\boldsymbol{\rho}$ & $\boldsymbol{n=33}$ & $\boldsymbol{n=100}$ \\
\hline\noalign{\smallskip}
$0$        & 33.0 & 100.0 \\
$2^{-100}$ & 33.0 & 99.0 \\
$2^{-64}$  & 33.0 & 64.0 \\
$2^{-33}$  & 32.0 & 33.0 \\
$2^{-20}$  & 20.0 & 20.0 \\
$2^{-10}$  & 10.0 & 10.0 \\
$2^{-3}$   & 3.0  & 3.0 \\
$1$        & 0.0  & 0.0 \\
\hline
\end{tabular}
\end{table}

\subsection{Common Causes}

\begin{figure}[t]
\centering
\includegraphics[width=\columnwidth]{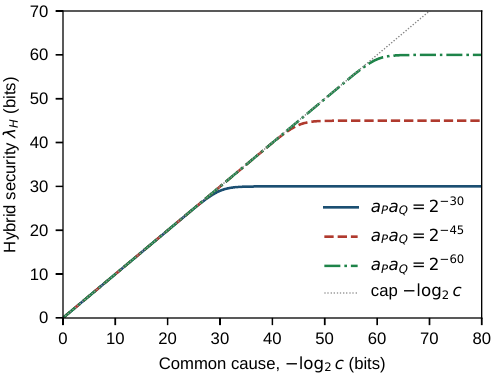}
\caption{Hybrid security $\lambda_H$ against the shared-cause probability $-\log_2 c$ for $a_Pa_Q=2^{-30}$ (solid), $2^{-45}$ (dashed) and $2^{-60}$ (dash-dot). The thin dotted diagonal is the cap $\lambda_H\le-\log_2 c$ of Theorem~\ref{thm:cc}}
\label{fig:cc}
\end{figure}

Table~\ref{tab:cc} lists values for Theorem~\ref{thm:cc} with $a_P=2^{-10}$, the estimated probability of flaws in the design or implementation of the post-quantum component over a given time period, and $a_Q=2^{-20}$, the estimated probability of flaws in the design or implementation of the QKD component. In the absence of shared design flaws, 30 bits of security are provided by the hybrid system, 20 more bits than by the post-quantum-only system. A shared design flaw with probability $2^{-20}$ cuts the security of the hybrid system in half, and a shared design flaw with probability $2^{-10}$, the same as $a_P$, adds just 1 bit of security to that provided by the post-quantum-only system. All values listed in each row of Table~\ref{tab:cc} are less than $\log_2(1+a_P/c)$. Figure~\ref{fig:cc} plots $\lambda_H$ versus $-\log_2 c$ to show that the security of the hybrid system is bounded below by that provided under the design-flaw assumptions and above by $-\log_2 c$.

\begin{table}[t]
\caption{Common-cause compromise (Theorem~\ref{thm:cc}) with $a_P=2^{-10}$, $a_Q=2^{-20}$: hybrid security $\lambda_H$, margins over standalone PQC ($\Delta_P$) and QKD ($\Delta_Q$), and the bound $\log_2(1+a_P/c)$ on $\Delta_P$, all in bits\textsuperscript{a}}
\label{tab:cc}
\centering
\small
\setlength{\tabcolsep}{4pt}
\begin{tabular}{@{}lrrrr@{}}
\hline\noalign{\smallskip}
$\boldsymbol{c}$ & $\boldsymbol{\lambda_H}$ & $\boldsymbol{\Delta_P}$ & $\boldsymbol{\Delta_Q}$ & \textbf{Bound} \\
\hline\noalign{\smallskip}
$0$        & 30.0 & 20.0 & 10.0 & -- \\
$2^{-60}$  & 30.0 & 20.0 & 10.0 & 50.0 \\
$2^{-40}$  & 30.0 & 20.0 & 10.0 & 30.0 \\
$2^{-30}$  & 29.0 & 19.0 & 9.0  & 20.0 \\
$2^{-20}$  & 20.0 & 10.0 & 1.0  & 10.0 \\
$2^{-10}$  & 10.0 & 1.0  & 0.0  & 1.0 \\
\hline
\end{tabular}

\smallskip
\raggedright\footnotesize\textsuperscript{a}Probabilities are stipulated for illustration, not measured
\end{table}

\subsection{Three Architectures}

Table~\ref{tab:arch} shows an application of Corollary~\ref{cor:multi} to three architectures with the same $a_P$ and $a_Q$ values. In architecture A, different keys are used for layer authentication, and host and operator domains are shared. In architecture B, an additional key is used for QKD post-processing authentication to remove the \texttt{auth} cause. In architecture C, host and operator domains for the subsystems are not shared, and only small domain overlaps remain. Reducing the largest domain overlap, from architecture A to B, reduces the PQC plus QKD security margin relative to standalone PQC by less than two bits because the next-largest domain overlap is with the operator domain. Architecture C, in which no large domain overlaps exist relative to $a_P$, offers the largest security margin.

\begin{table}[t]
\caption{Shared-cause decomposition $c=1-\prod_i(1-c_i)$ for three architectures, with $a_P=2^{-10}$, $a_Q=2^{-20}$. Entries for $c_i$ are $\log_2$ probabilities\textsuperscript{a}}
\label{tab:arch}
\centering
\small
\setlength{\tabcolsep}{2.5pt}
\begin{tabular}{@{}lcccrrr@{}}
\hline\noalign{\smallskip}
\textbf{Arch.} & \texttt{auth} & \texttt{host} & \texttt{op} & $\boldsymbol{-\log_2 c}$ & $\boldsymbol{\lambda_H}$ & $\boldsymbol{\Delta_P}$ \\
\hline\noalign{\smallskip}
A: shared      & $-12$ & $-16$ & $-14$ & 11.6 & 11.6 & 2.0 \\
B: sep.\ auth  & --    & $-16$ & $-14$ & 13.7 & 13.7 & 3.8 \\
C: sep.\ domains & --  & $-40$ & $-36$ & 35.9 & 30.0 & 20.0 \\
\hline
\end{tabular}

\smallskip
\raggedright\footnotesize\textsuperscript{a}Probabilities are stipulated for illustration, not measured; ``--'' denotes a removed cause
\end{table}

\section{Design Implications and Limitations}
\label{sec:criteria}

The results yield three criteria for hybrid deployment.

\begin{criterion}[Common causes first]
\label{crit:cc}
Adding QKD to a post-quantum system offers no more than $\log_2(1+a_P/c)$ bits of security (Theorem~\ref{thm:cc}). To justify the use of a quantum channel over a post-quantum system, it will generally be necessary to address other shared vulnerabilities that could compromise the post-quantum system with high probability (Corollary~\ref{cor:multi}). Possible solutions include separate authentication keys or pre-shared keys for QKD operations, separate hosts, and separation of administrative domains for QKD operations.
\end{criterion}

\begin{criterion}[Sound combination without silent fallback]
\label{crit:sound}
Use the equations in \eqref{eq:ser} to \eqref{eq:salt}, where $\ctx$ includes identifiers for the KEM ciphertext and QKD key (Proposition~\ref{prop:modes}). Do not include a fall-back option to use post-quantum keys instead of QKD if QKD fails (Proposition~\ref{prop:fallback}). This recommendation conflicts with standard engineering practice regarding fall-back protections and merits explanation. Not only does failure of a QKD session represent an accident, but it also signals an active attack on the quantum channel, an attacker who blocks or corrupts quantum transmissions at a rate exceeding the QKD abort threshold can cause QKD to fail. Fall-back protections guard against accidents; active attackers can choose when to trigger fall-back protections. A silent fall-back option enables an attacker to force the use of only post-quantum keys in all sessions by blocking quantum transmissions without other negative effects on other aspects of the application protocol. There are no signs of attack on the application protocol. Loss of the QKD warning signal reduces the effective security of all sessions to  $\lambda_P$. If a pure post-quantum mode is required for availability, it can be offered as a limited service with usage logs to alert users and operators to a potential attack, but not as a silent default. Using the QKD key directly is not a hybrid protocol; it is QKD plus post-quantum authentication (Remark~\ref{rem:direct}).
\end{criterion}

\begin{criterion}[Quantum floor]
\label{crit:floor}
The scheme is provably secure against total compromise of the post-quantum assumption at level $2^{-\lambda_T}$ if and only if $\lambda_Q\ge\lambda_T$ (Proposition~\ref{prop:degrade}). The value $\lambda_Q$ depends on device and authentication attacks and other typical attacks and can be verified up to $-\log_2(q+\varepsilon_Q)$. The main advantage of hybrid schemes is enhanced long-term security. This has important deployment considerations for physical-layer engineers to keep in mind about QKD: The security provided by QKD systems is limited to the level provided by finite-key analysis. With $\varepsilon_Q=10^{-10}\approx2^{-33}$ per key block, even authenticated, trusted devices can provide only 33 or fewer bits of quantum security. In a typical hybrid quantum-classical network, connecting to a 128-bit PQC system will not provide 128 bits of information-theoretic security against PQC failure; 128 bits of quantum security would require $\varepsilon_Q\le2^{-128}$ and reduced key generation rates.
\end{criterion}

\paragraph{Limitations.}
First, Assumption~\ref{asm:exo} excludes attacks that exploit knowledge of key values; see Backendal et al.'s \cite{backendal2025kdf} work for attacks of this type. Second, the all-or-nothing by layer compromise assumption makes the entire key for a layer known if that layer is compromised. This is a conservative assumption that prevents us from comparing combination strategies under partial compromise. Third, the $\mathcal{D}$ distributions are modeling parameters. We include $\mathcal{D}$ explicitly in the model and bound effects of these parameters but do not provide tools to help with choosing realistic values of $\mathcal{D}$ for deployment. Fourth, we model a single-session interaction between two parties. Trusted intermediaries do not directly participate in our models but may interact with sub-components that reveal information about $Q$, or both layers, if collocated with the PQC component. Measurement-device-independent QKD \cite{measurement-device-independent_lo_2012} is a special case of the \texttt{qkd} component.

\section{Conclusion}

Combination theorems show that a combined PQC-QKD secret key is at least as secure as its stronger component. We present results showing circumstances under which a combined secret key is more secure than its stronger component. In an indistinguishability experiment with exogenous compromise, the best attack on a combined secret key has a success probability of the product of the success probabilities of attacks on individual key layers, ignoring terms near zero. A combined secret key's advantage over its stronger component is $\min(n,-\log_2\rho)$, plus or minus 1 bit. Common factors affecting key security reduce the security of a combined secret key to $-\log_2 c$ bits and a combined secret key's advantage over a pure PQC secret key to $\log_2(1+a_P/c)$ bits. Using a fallback to a pure PQC secret key in case of QKD failure eliminates the advantage of a combined secret key. Benefits of using a quantum channel include reductions in common factors affecting key security, such as shared authentication equipment and staff, compared with benefits of using pure PQC or QKD secret keys.

Possible future directions of research are using the model to estimate the compromise distribution based on observation of compromise incidents and extending the model to handle partial and relayed compromises. Direct study of shared causes discussed here is also a possibility.

\section*{Statements and Declarations}

\noindent\textbf{Funding.} This work was supported by Qatar University.

\medskip\noindent\textbf{Competing interests.} The authors declare no competing interests.

\medskip\noindent\textbf{Ethics approval.} Not applicable. The study involves no human participants, animals or personal data.

\medskip\noindent\textbf{Consent to participate and consent for publication.} Not applicable.

\medskip\noindent\textbf{Data availability.} No datasets were generated or analyzed during the current study. All numerical values are evaluated from closed-form expressions given in the paper.

\medskip\noindent\textbf{Code availability.} The script that computes Tables~\ref{tab:margin}--\ref{tab:arch} and Figs.~\ref{fig:margin}--\ref{fig:cc}, and checks the stated bounds numerically, is provided as Online Resource~1.

\medskip\noindent\textbf{Author contributions.} All authors contributed to the conception and design of the study. The first draft of the manuscript was written by Mona Alotaibi, and all authors commented on previous versions. All authors read and approved the final manuscript.

\appendix
\section{Proofs}
\label{app:proofs}

\paragraph{Proof of Theorem~\ref{thm:bridge}.}
\emph{Upper bound.} Under Assumption~\ref{asm:exo}, the random variable $S$ is independent of $b$ and the layers' randomness.
\[
\Adv(\mathcal{A})\le\sum_{s}\Pr[S=s]\;\Adv_s(\mathcal{A}).
\]
where $\Adv_s$ is the advantage in the experiment with string $S=s$ fixed. We bound $\Adv_s$ according to the exposure of $s$. For an individual $s$ with both layers fully exposed, $\Adv_s\le1$.

Suppose $P$ is not shielded. Let $G_0$ be a game in which $S=s$, and let $G_1$ be a game like $G_0$ except that all occurrences of $K_P$ in the computation of $K$ are replaced by uniformly random independent copies $U_P$. A distinguisher that, on input a PQC transcript and a string $Y\in\{K_P,U_P\}$, uses the input to simulate the honest QKD layer (whose output is independent of $K_P$), runs $\mathcal{A}$ with access to the simulated layer, computes $K$ from $Y$, and continues the game has output distributions matching those of $G_0$ or $G_1$ and auxiliary input distributions independent of $K_P$ given a PQC transcript; by Assumption~\ref{asm:comp}(a), we have $|\Pr_{G_0}[b'=1\mid b=0]-\Pr_{G_1}[b'=1\mid b=0]|\le\delta_P(t')$. For $b=1$, the distinguisher's view is independent of $K$ and the same as its view in $G_0$. For $b=0$, $\Comb(U_P,K_Q,\ctx)$ with $K_Q$ and $\ctx$ known to the simulator is $\delta_{c,P}(t')$-indistinguishable from uniform by Assumption~\ref{asm:comp}(c); replacing $\Comb(U_P,K_Q,\ctx)$ with an independent uniformly random string gives a game where $b=0$ and $b=1$ are identically distributed. Thus, $\Adv_s\le\eta_P$.

Suppose that $Q$ is unknown. Then $\texttt{auth}\notin s$. The classical post-processing channel is authenticated, and Assumption~\ref{asm:comp}(b) is satisfied. Using the ideal state to simulate other systems, including $K_Q$, can change the probability of any event observed by a quantum distinguisher by at most $\varepsilon_Q$, since other steps in the process are completely positive and trace-preserving. In the ideal state, $K_Q$ is uniformly distributed and independent of all other systems, including $K_P$, and Assumption~\ref{asm:comp}(c) is satisfied for an adversary knowing $K_P$. The bound follows as before. Thus $\Adv_s\le\eta_Q$.

If both layers are latent, both terms in the equation are present. Summing over patterns gives \eqref{eq:bridge}, and the second equality comes from the fact that $\pi_{10}+\pi_{01}+\pi_{00}\le1$.

\emph{Lower bound.}$\mathcal{A}^*$is passive. On inputs both keys, $\mathcal{A}^*$ outputs 1 precisely for inputs $K_b=\Comb(K_P,K_Q,\ctx)$ and 0 otherwise. On sessions where both layers are eventually revealed, $\mathcal{A}^*$ outputs 1 with probability 1 if $b=0$ and $2^{-\ell}$ if $b=1$, and its output does not otherwise depend on $b$. Because $S$ is independent of the honest abort event, $\Adv(\mathcal{A}^*)=(1-p_\bot)\pi_{11}(1-2^{-\ell})$. 

\paragraph{Proof of Corollary~\ref{cor:exact}.}
By \eqref{eq:bridge}, we have $\Adv^*\le\pi_{11}+\max(\eta_P,\eta_Q)\le2\pi_{11}$, or $-\log_2\Adv^*\ge\lambda_H-1$. The lower bound in Theorem~\ref{thm:bridge} gives the opposite inequality. A similar argument applies to a one-layer system. 

\paragraph{Proof of Proposition~\ref{prop:range}.}
Under fixed marginals, the joint distribution of $(\mathbf{1}_{E_P},\mathbf{1}_{E_Q})$ depends only on $\pi_{11}$, and any $\pi_{11}\in[\max(0,p+q-1),\min(p,q)]$ is possible by setting $\pi_{10}=p-\pi_{11}$, $\pi_{01}=q-\pi_{11}$, $\pi_{00}=1-p-q+\pi_{11}$, all nonnegative. If $p+q\le1$, then $\pi_{11}$ can take on a minimum value of 0. The correlation $\rho$ is a linear increasing function of $\pi_{11}$, and $\rho=1$ when $\pi_{11}=\min(p,q)$, $\rho=0$ when $\pi_{11}=pq$, and, noting that $\min(p,q)-pq=\min(p,q)(1-\max(p,q))$, $\rho=-\max(p,q)/(1-\max(p,q))$ when $\pi_{11}=0$. 

\paragraph{Proof of Proposition~\ref{prop:margin}.}
From Definition~\ref{def:rho}, $\pi_{11}/\min(p,q)=(1-\rho)\max(p,q)+\rho=2^{-n}+\rho(1-2^{-n})$, as shown in \eqref{eq:margin}. Let $B$ be the bracket. If $\rho\in[0,1]$, then $B\ge2^{-n}$ and $B\ge\rho$, so $\Delta\le\min(n,-\log_2\rho)$. If $\rho\in[0,1]$, then $B\le2^{-n}+\rho\le2\max(2^{-n},\rho)$, so $\Delta\ge\min(n,-\log_2\rho)-1$. If $\rho<0$, then $B<2^{-n}$, so $\Delta>n$. 

\paragraph{Proof of Proposition~\ref{prop:floor}.}
$\pi_{11}\le\min(p,q)$ implies $\lambda_H\ge m$. If $p=1,$ then event $E_P$ occurs almost surely and $\pi_{11}=q$. 

\paragraph{Proof of Theorem~\ref{thm:cc}.}
$E_P\cap E_Q=C\cup(A_P\cap A_Q)$, and independence gives $\pi_{11}=c+(1-c)a_Pa_Q$; there is also $p=a_P+c-a_Pc$. (i) The condition $\pi_{11}\ge c$ gives an upper bound on $\lambda_H$, and $\pi_{11}\le c+a_Pa_Q\le2\max(c,a_Pa_Q)$ gives a lower bound. (ii) $\Delta_P=\log_2(p/\pi_{11})$ and
\[
\frac{p}{\pi_{11}}=\frac{c+(1-c)a_P}{c+(1-c)a_Pa_Q}\le\frac{c+(1-c)a_P}{c}\le1+\frac{a_P}{c}.
\]
The upper bound on $\Delta_Q$ is symmetric. For $c\ge a_P$, the upper bound on $\Delta_Q$ is no more than $\log_2 2=1$. 

\paragraph{Proof of Corollary~\ref{cor:multi}.}
$C$ is true if any of the $C_i$ are true. So $c=1-\prod_i(1-c_i)$. We have $c\ge\max_ic_i$, since $C$ contains $C_i$, and $c\le\sum_ic_i$, by the union bound. Part (i) of Theorem~\ref{thm:cc} then gives $\lambda_H\le-\log_2c\le-\log_2\max_ic_i$. Removing all but one cause replaces $c$ by $\max_ic_i\ge c/k$, and increases $-\log_2c$ by no more than $\log_2k$. 

\paragraph{Proof of Proposition~\ref{prop:degrade}.}
$\pi_{11}=c+(1-c)a_Pa_Q$ is nondecreasing in $a_P$. The constraint $\pi_{11}\le2^{-\lambda_T}$ can be rewritten $(1-c)a_Pa_Q\le2^{-\lambda_T}-c$. For this inequality to hold, it is necessary that $c\le2^{-\lambda_T}$ and, assuming $c<1$ and $a_Q>0$, sufficient that $a_P\le a_P^*$. If $a_P=1$, then $\pi_{11}=c+(1-c)a_Q=q$. 

\paragraph{Proof of Proposition~\ref{prop:modes}.}
In each case, we prove that we meet Assumption~\ref{asm:comp}(c): HKDF-Expand is a PRF keyed by the extracted value \cite{krawczyk2010hkdf}. For \eqref{eq:ser}, the extracted value is $\mathrm{HMAC}(0,K_P\|K_Q)$, which is pseudorandom when one of the inputs is uniformly random and secret and the other is known, as shown by the dual-PRF (message-position) property of HMAC \cite{backendal2023hmac} or the multi-input KDF definition of \cite{backendal2025kdf}. For \eqref{eq:par}, if $K_P$ is uniformly random and secret, then the first input is pseudorandom by HKDF security with a public salt \cite{krawczyk2010hkdf}. The second input depends only on known inputs $k_Q$ and $\ctx$ and is independent of $K_P$. The XOR of a pseudorandom string with an independent known string is pseudorandom. The proof for $K_Q$ is similar. For \eqref{eq:salt}, $\mathrm{HKDF\text{-}Extract}(K_P,K_Q)=\mathrm{HMAC}(K_P,K_Q)$. If $K_P$ is uniformly random and secret, then this is pseudorandom for any known $K_Q$ due to HMAC's PRF security in its key. If $K_Q$ is uniformly random and secret and $K_P$ is known, then this matches the extraction setting of \cite{krawczyk2010hkdf} with a known salt.

\paragraph{Proof of Proposition~\ref{prop:fallback}.}
$\mathcal{A}$ blocks all quantum communication through the channel, so it receives no photons, and QKD always aborts. Otherwise, $\mathcal{A}$ acts like $\mathcal{B}$, forwarding all inputs and outputs. For QKD aborts, $K$ computed by $\Pi_{\mathsf{fb}}$ is the same as $K$ computed by $\Pi_P$. Actions by the PQC layer, use of $K_P$, and disclosure of $K_P$ are the same, and information about $K_P$ is not disclosed in QKD communications. The view of $\mathcal{B}$ in the experiment for $\Pi_{\mathsf{fb}}$ is statistically equivalent to its view in the experiment for $\Pi_P$, and the advantages are equal. The final result follows from applying the lower-bound argument used in Theorem~\ref{thm:bridge} to $\Pi_P$, with the success probability of the PQC layer replacing $1-p_\bot$.

\end{document}